\documentclass[final]{aipproc}
\layoutstyle{8x11double}

\usepackage{xspace}

\newcommand{\png}{PN\,G135.9+55.9\xspace}
\newcommand{\sbs}{SBS\,1150+599A\xspace}

\newcommand{\teff}{\ensuremath{T_{\mathrm{eff}}}}

\begin{document}

\title{On the most metal-poor PN and its binary central 
star\footnotemark[1] \footnotemark[2] \footnotemark[3] \footnotemark[4]
 \footnotemark[5]
}



\author{R.~Napiwotzki}{
  address={Centre for Astrophysics Research, STRI, University of
  Hertfordshire, College Lane, Hatfield AL10 9AB, UK}
}

\author{G.~Tovmassian}{
  address={Observatorio Astron\'omico, Instituto de Astronom\'ia,
  UNAM, P.O.\ Box 439027, San Diego, CA 92143-9027}
}

\author{M.G.~Richer}{
  address={Observatorio Astron\'omico, Instituto de Astronom\'ia,
  UNAM, P.O.\ Box 439027, San Diego, CA 92143-9027}
}

\author{G.~Stasi\'nska}{
  address={LUTH, Observatoire de Meudon, 5 Place Jules Janssen,
  F-92195 Meudon Cedex, France}
}

\author{M.~Pe\~na}{
  address={Instituto de Astronom\'ia, UNAM, Apdo.\ Postal 70 264, 
  M\'exico D.F., 04510, M\'exico}
}

\author{H.~Drechsel}{
  address={Dr.~Remeis-Sternwarte, Sternwarstr.~7, 96049 Bamberg, Germany}
}
\author{S.~Dreizler}{
  address={Institut f\"ur Astrophysik, Universit\"at G\"ottingen, 
  Friedrich-Hund-Platz 1, 37077 G\"ottingen, Germany}
}
\author{T.~Rauch}{
  address={Inst.\ f\"ur Astronomie und Astrophysik T\"ubingen (IAAT), 
  Sand 1, 72076 T\"ubingen, Germany}
}

\begin{abstract}
\png is the most metal-poor PN known in our Galaxy. The
central star resides in a short-period binary system with a compact
component, probably a white dwarf. We describe new observations, which
allowed us to determine the orbital period. The lower limit for the
combined mass of both stars is close to the Chandrasekhar limit for
white dwarfs, making this binary a possible progenitor of a supernova
type Ia. The binary system must have recently emerged from a common
envelope phase. 
\end{abstract}

\maketitle

\footnotetext[1]{Based on 
observations obtained at the Canada-France-Hawaii Telescope (CFHT)
which is operated by the National Research Council of Canada, the
Institut National des Science de l'Univers of the Centre National de
la Recherche Scientifique of France, and the University of Hawaii}

\footnotetext[2]{Based on observations collected at the Centro Astron\'omico 
Hispano Alem\'an (CAHA) at Calar Alto, operated jointly by the Max-Planck
Institut f\"ur Astronomie and the Instituto de Astrof\'isica de Andaluc\'ia
(CSIC)}

\footnotetext[3]{Based upon observations obtained at the Observatorio 
Astron\'omico Nacional in Sierra San Pedro M\'artir, Baja California, Mexico}

\footnotetext[4]{Based on observations obtained with the Hobby-Eberly 
Telescope, which is a joint project of the University of Texas at
Austin, the Pennsylvania State University, Stanford University, 
Ludwig-Maximilians-Universit\"at M\"unchen, and
Georg-August-Universit\"at G\"ottingen}

\footnotetext[5]{Based
on observations made with the NASA/ESA Hubble Space
Telescope, obtained from the Data Archive at the Space Telescope
Science Institute, which is operated by the Association of
Universities for Research in Astronomy, Inc., under NASA contract NAS
5-26555. These observations are associated with program \# 9466}

%%%%%%%%%%%%%%%%%%%%%%%%%%%%%%%%%%%%%%%%%%%%
%% MAINMATTER
%%%%%%%%%%%%%%%%%%%%%%%%%%%%%%%%%%%%%%%%%%%%

\section{Introduction}

\begin{figure}
\resizebox{0.95\columnwidth}{!}
   {\includegraphics[angle=0]{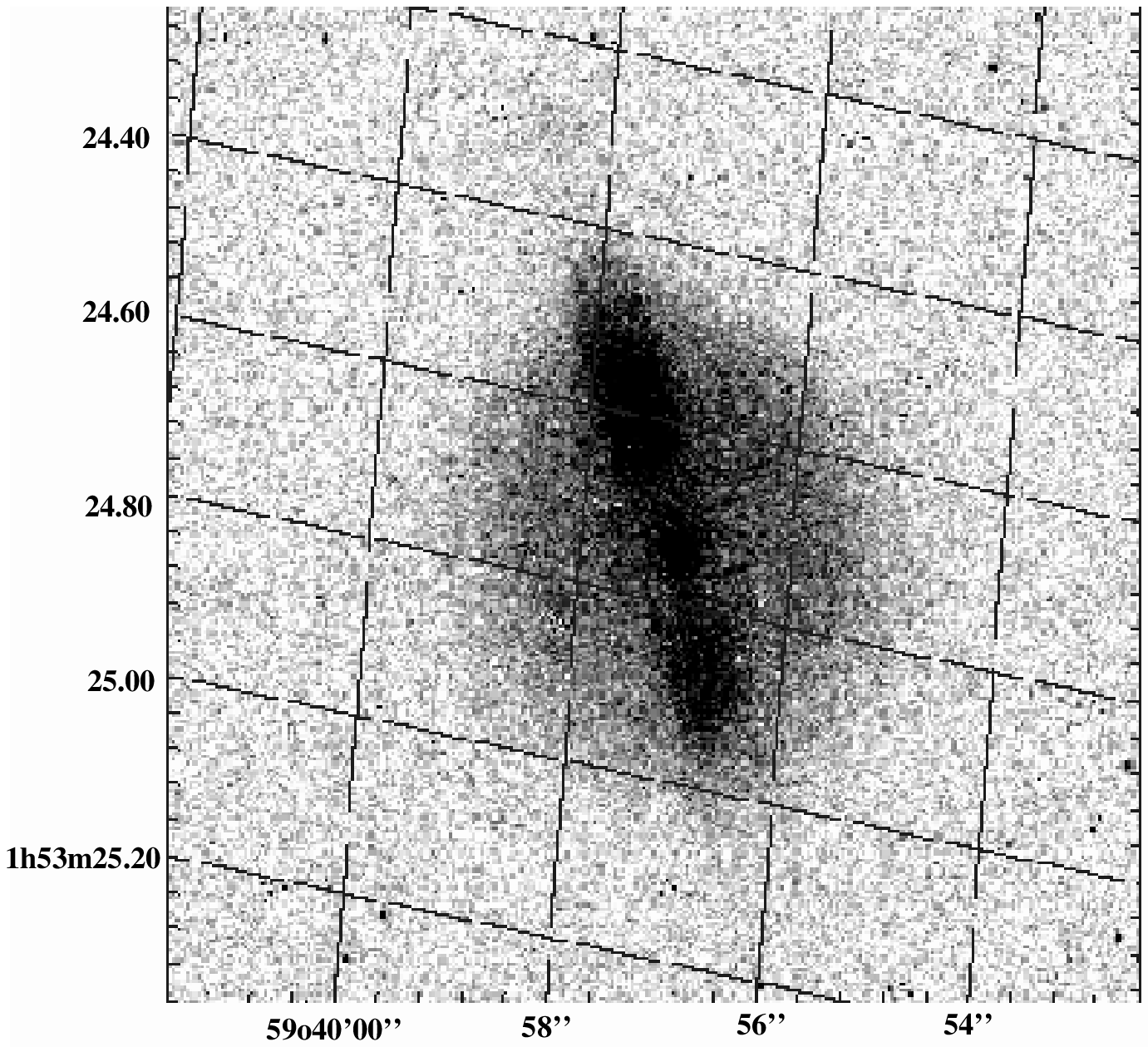}}
  \caption{HST H$\alpha$ image of \sbs.}
% with overlayed contour lines.}
  \label{f:HST}
\end{figure}

\png, also known as \sbs, is a planetary nebula (PN) discovered in the Second
Byurakan Sky Survey. 
% was originally discovered by the Second Byurakan Sky Survey and
% listed under the name \sbs as cataclysmic variable.  Its real nature
% was uncovered by \citet{TSC2001} who showed that \sbs is a planetary
% nebula (PN).  
The spectrum of \sbs is quite unusual for a planetary
nebula, showing only lines of H\,I, He\,II and extremely weak
forbidden lines of heavier elements \citep{TSC2001}.
% , which was the likely reason
% for the initial mis-classification. 
A recent HST/STIS image shows \png
as a PN with axial-symmetric density enhancement (Fig.~\ref{f:HST}).

Initial analyses with photoionisation models indicated
that the oxygen abundance is very low, around 1/100 of the solar value
\citep{TSC2001}.
%,RTS2002,JFC2002} with an upper limit of 1/13 solar 
%\citep{PT2005}. However, all these analyses suffered from the lack of
%information on the abundances of carbon and nitrogen, which do not
% produce detectable lines in the optical spectrum of \png. These
Carbon and nitrogen, which do not
produce detectable lines in the optical spectrum of \png, 
have an important impact on the thermal balance of the PN
and, indirectly, on the calculated oxygen abundance.  New measurements
of crucial line intensities in the UV by the HST-STIS spectrograph
allow to put more stringent constrains on the carbon and
nitrogen abundances, and thus on the nebular oxygen abundance.
% \citet{STR2005} presented a new abundance analysis incorporating
The oxygen abundance is now constrained between
1/125 and 1/40 of the solar value \citep{STR2005}, 
confirming \png as the most oxygen poor PN known.  

The low oxygen abundance indicates a population~II nature of
\png. Further evidence comes from a very large distance from the Galactic
plane \citep[15\,kpc;][]{TNR2004} and a very high heliocentric radial
velocity \citep[$-193$\,km/s;][]{RLS2003}.

\begin{figure}
\resizebox{0.9\columnwidth}{!}
   {\includegraphics[angle=-90]{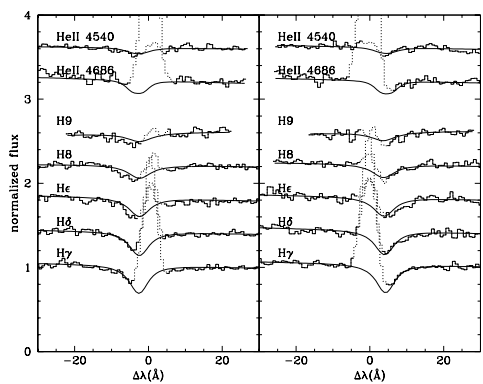}}
  \caption{Model atmosphere fit to two optical CFHT spectra
  taken close to the phases of largest RV shift. The dotted lines
  indicate parts of the spectra excluded from the fit.}
  \label{f:balmer}
\end{figure}

In May 2003 we obtained a series of medium resolution spectra of \png
with the 3.6\,m CFHT \citep{TNR2004}. 
% The main aim were accurate
% measurements of the [Ne\,V]~3426\,\AA\ and [Ne\,III]~3869\,\AA\ nebula
% lines. As a by-product we aquired 
This produced a good S/N spectrum of the central
star (CSPN). Photospheric hydrogen Balmer lines were clearly
visible. Closer inspection revealed that the wavelength of the Balmer
lines varied from exposure to exposure, indicating large radial
velocity (RV) variations caused by orbital motion in a close binary
(see Fig.~\ref{f:balmer}).
Since the observations were not optimised for the determination of
orbital periods, it was not possible to determine an unambiguous
period. However, we could show that 4\,h was an upper limit with
candidate values at 0.94\,h, 1.5\,h and 3.6\,h. 

The stellar temperature is constrained by the CSPN spectrum and the
photoionisation models to 100,000\ldots 120,000\,K. We adopt a CSPN
mass of $0.55M_\odot$ \citep[cf.\ discussion in ][]{TNR2004}.

\section{New observations}

In April 2004 we secured a short series of photometric observations of
\png in the $V$ band
with the 2.1\,m telescope at San Pedro M\'artir/Mexico (SPM). 
The brightness of
the CSPN was measured differentially to comparison stars. These
observations revealed a clear photometric variability on the 0.1\,mag
level. However, the run was too short to determine its period and 
the precise nature of the variability.

Two short series covering one hour each of medium resolution spectra
were obtained in May 2004 with the Hobby-Eberly telescope of McDonald
observatory. These new RV measurements supported our longest candidate
period.

In January 2005 we ran a campaign at three observatories to get
near-simultaneous photometric and spectroscopic
observations. Two half-nights of spectroscopy were scheduled at the
Keck~I telescope. Three nights of photometry were granted for the
2.2\,m telescope of the Calar Alto observatory/Spain equipped with the
BUSCA camera and one night at
the SPM observatory. BUSCA is a CCD camera which is equipped
with dichroic beam splitters to allow simultaneous
observations in four colour bands.
%  \citep{RPG2000}. 
We used BUSCA without additional filters to be able to get a good time
resolution with the best possible S/N. This resulted in four
instrument specific broadband channels: UV (approximately 3500\ldots
4200\AA), blue (4200\ldots 5100\,\AA), red (5100\ldots 7200\,\AA) and
IR (7200\ldots 9000\,\AA). SPM observations were carried out with a
$V$ filter.
%The chosen exposure time was 120\,s, which
% resulted in a duty cycle of 280\,s. \remind{SPM}

While weather conditions at Calar Alto and SPM were good and both runs
produced valuable results, poor weather conditions prevented us
from observing at the Keck telescope. 
The brightness of the CSPN was measured differentially to comparison stars in
the field. 
% Colour terms were negligible with the exception of the UV
% band. For the latter band we performed a correction for the airmass
% dependent differential extinction.

%% Care was taken to select suitable blue reference stars to minimize
%% color effects. Since ven the bluest stars in the field are much redder than
%% \png we checked for dependencies of the differential measurements on
%% airmass. It turned out that airmass caused only negligible effects in most
%% filters, with the exception of the UV band. Thus measurements in the latter
%% band were airmass corrected before analysing the light curve.

\subsection{The light curve and its interpretation}

\begin{figure}
\resizebox{0.9\columnwidth}{!}
   {\includegraphics{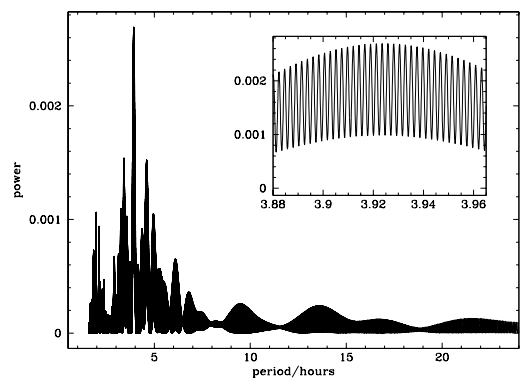}}
  \caption{Power spectrum of the three photometric runs combined.}
  \label{f:pow}
\end{figure}

\begin{figure}
\resizebox{0.9\columnwidth}{!}
   {\includegraphics[angle=-90]{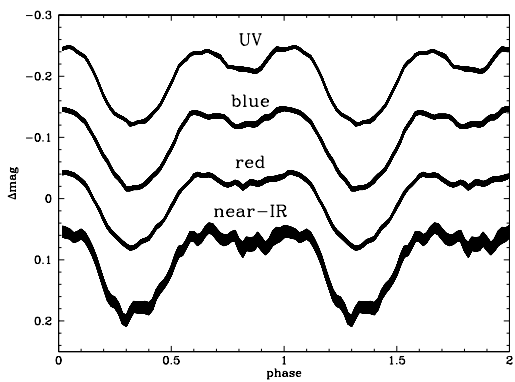}}
  \caption{Phase folded photometry for the four broadband filters of the 
  BUSCA camera taken during the Calar Alto run in Jan.\ 2005. The widths
  of the curve indicate the error limits.}
  \label{f:BUSCA}
\end{figure}

We combined the data from all three photometric runs (April 2004 and Jan.\
2005) for the period determination. The resulting power spectrum is shown in
Fig.~\ref{f:pow}. It is obvious that the period ambiguity present in the CFHT
radial velocity data has now been overcome. The power spectrum shows a clear
peak at 3.92\,h, consistent with the longest candidate period derived from the
spectroscopic measurements. Fig.~\ref{f:BUSCA} shows the BUSCA four channel
photometry, folded with the photometric period.

Inspection of the inset in Fig.~\ref{f:pow} reveals that alias peaks
of comparable strength are present. These are separated by 8.5\,s, which
corresponds to one cycle over the nine months time gap between the April 2004
and the Jan.\ 2005 runs. These aliases are of negligible importance for the
interpretation of this system, but prevent us from computing reliable
photometric phases for the spectra taken with the CFHT in May 2003.
The new accurate period combined with the RV curve enables us to
compute the mass function and a lower limit of the mass of the
companion: $M_{\mathrm{comp}} = 0.85M_\odot$.

A few CSPN in close binaries with a main sequence companion are known
to show photometric variability. A well investigated example is BE\,UMa.
% \citep{CCH1983,FJ1994,WRZ1995,LTN1995,FLH1999}. 
This system consists of a $\approx$105,000\,K hot CSPN, i.e.\ similar
to the CSPN in \png, and a K dwarf companion
\citep{LTN1995,FJ1994,WRZ1995}, which is over-sized for its mass of
$0.36M_\odot$ \citep{FLH1999}.  The BE\,UMa system is much wider than
the \png binary. Its orbital period is 2.29\,days.

BE\,UMa shows a dramatic reflection effect causing a photometric
amplitude of 1\,mag \citep{Kur1971}. The strong radiation of the hot
CSPN is reprocessed on the illuminated side of the main sequence
companion. The luminosity of the system depends on the visible
fraction of the illuminated hemisphere, which varies with the orbital
phase. The EUV radiation of the very hot CSPN is responsible for a
forest of recombination lines in emission \citep[see, e.g., Fig.~1 in ][]
{FJ1994}. During most of the orbit the CSPN spectrum is virtually
swamped by the companion's emission line forest and special care has
to be taken to secure a relatively uncontaminated spectrum of the CSPN
\citep{LTN1995}. 

Since no such forest of emission lines can be seen in our \png spectra
(Fig.~\ref{f:balmer}), we concluded that the companion in this system
must be a compact object, i.e.\ a white dwarf or a neutron star
\citep{TNR2004}. \citet{PT2005} objected that the \png system is more
extreme than BE\,UMa. They speculated that the intense radiation of the
CSPN has the result that ``the photosphere of a 'main sequence'
companion could be so hot as to emit a spectrum totally different from
the one of BE UMa'' \citep{PT2005}.

We agree insofar as the ionisation balance is likely shifted from the
predominant lines of O\,II, N\,II, C\,II/III towards higher ionisation
stages, maybe as high as O\,VI, N\,V and C\,IV. However, these species
have optical lines, which should be visible. Moreover, the heating of
the parts of the illuminated hemisphere closer to the edge should be
less extreme, which should allow for the presence of lower ionisation
stages.  Nevertheless, we will explore whether our new photometric
light curve is compatible with a main sequence companion.

What effects can cause photometric variability in a close CSPN -- main
sequence star system? 
\begin{description}
\item[Reflection effect:]  This was already described above.
\item[Eclipses:] For the given dimensions of the system deep eclipses would be
  expected, unless the system inclination is below $i = 60^{\circ}$.
\item[Ellipsoidal variation:] Both, CSPN and even more so the hypothetical main
  sequence star, would fill a major fraction of their Roche lobe. The
  gravitational pull of the companion causes deviations from spherical
  symmetry -- an elongated ellipsoid in first approximation. The cross
  section and thus the brightness of the star(s) depends on their
  phase angle.
\end{description}
Note that the periodicity of the reflection effect is equal to the
orbital period, while it is twice that for ellipsoidal light variation.

The multi-colour light curves were analysed with the Wilson-Devinney
based solution code MORO \citep{DHL1995}, which is optimised for the
treatment of early-type close binaries by the inclusion of radiative
pressure effects. All relevant physical and geometrical parameters of
the system can be simultaneously adjusted.

A rock bottom lower limit for a main sequence star of the required
mass ($0.85M_\odot$) 
is $R=0.5R_\odot$ \citep{BCA1997}. It turned out to be impossible
to produce an acceptable fit to the observed light curve with a
companion of this size. Die-hard fans of the main sequence companion 
could now argue that the companion might be undersized, because it 
has just recovered from the common envelope event. This is not too far
fetched, because it is known that the main sequence star in the
BE\,UMa system is still over-sized for its mass \citep{FLH1999}. Thus
we treated the radius as a free parameter. In this case we were able
to find a reasonable fit of the light curve. The resulting radius is
$R=0.18R_\odot$ and the inclination angle $i=40^\circ$. However, 
the mass function derived from the RV curve requires 
$1.8M_\odot$ for this solution. This is far higher than any plausible
mass for a main sequence companion. Thus the result of the light curve
analysis supports our earlier conclusion \citep{TNR2004} that the
companion must be a (pre-?) white dwarf or a neutron star. 

\begin{figure}
\resizebox{1.0\columnwidth}{!}
   {\includegraphics[angle=-90,bb=290 16 596 784]{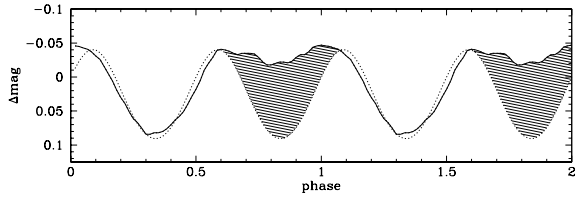}}
  \caption{The phase-folded ``blue''-light curve from
  Fig.~\ref{f:BUSCA} is compared to a sine-wave (dotted line), a first order
  approximation to the effect to ellipsoidal variability. The hatched
  region corresponds to the ``excess luminosity''.}
  \label{f:ellipsoidal}
\end{figure}

Even without the main sequence companion an ellipsoidal light
variation does not come unexpected, because the CSPN fills a
significant fraction of its Roche lobe \citep[cf.\ Table~2 and~3
in][]{TNR2004}. We compare the measured light curve in
Fig.~\ref{f:ellipsoidal} to the simplest possible ellipsoidal light
curve, a sine-wave. As explained above the period of ellipsoidal
variation is twice that of the orbital period. The first half of the
observed light curve is well explained, but in the second half we see
an excess over the predicted curve. 

What can cause this extra light? One possible explanation could be a
hot spot on the CSPN caused by \mbox{(particle-)}irradiation from a neutron
star companion. Alternatively, the extra light could stem from a
tilted disk-like structure around the white dwarf/neutron star
companion. This could possibly result from accretion from the CSPN wind 
or be a remnant from the recent common envelope phase. None of these
models is without problems. Relative phasing of the binary
orbit and the photometric variability will allow to distinguish
between both possible models of the system. 

\section{Summary and outlook}

\sbs is the hot ($\teff \approx
120,000$\,K) central star of a very metal poor PN. The CSPN resides in
a close binary system with a period $P=3.92$\,h as determined from the
photometric light curve. The minimum companion mass is
$M_2=0.85M_\odot$. Since a main sequence component can be ruled out,
the second star must be a compact object, most likely a white dwarf or
alternatively a neutron star. Gravitational waves will cause the
merging of the binary in 1\,Gyr. Since the minimum mass of the system
$M\ge 0.55M_\odot +0.85 M_\odot$ is close to the Chandrasekhar limit,
\sbs is a very good SN\,Ia progenitor candidate within the double
degenerate scenario. 

\citet{NCD2003} conducted an RV survey (SPY) for
DD progenitors of SN\,Ia as a large programme at the 8\,m ESO
VLT. About 100 DDs were detected, including only one good SN\,Ia progenitor
candidate. This demonstrates the rareness of these objects. Had it not
been for the PN and the extraordinary chemistry of \png the DD central
star would have remained unnoticed. The SPY sample will provide
information like period and mass distribution of DDs. However, the
CSPN in \png can be considered as DD in statu nascendi. The
investigation of the PN offers unique opportunities to constrain the
poorly understood common envelope phase. The kinematics of the PN
constrain the dynamics and timescales of the CE event. The PN
chemistry enables us to evaluate the dredge-up of processed material.

% Crucial for a better understanding of the
% \sbs system will be spectroscopic and photometric observations taken
% close enough in time to determine the relative phasing of the binary
% orbit and the photometric variability. This will allow to distinguish
% between the possible models of the system. When this information is
% available a combined analysis will help to constrain the inclination
% of the orbit, the radius of the CSPN and eventually fix the mass of
% the companion.

%%%%%%%%%%%%%%%%%%%%%%%%%%%%%%%%%%%%%%%%%%%%%%%%
%% BACKMATTER
%%%%%%%%%%%%%%%%%%%%%%%%%%%%%%%%%%%%%%%%%%%%%%%%

% \begin{theacknowledgments}
% R.N.\ acknowledges support by a PPARC Advanced Fellowship.
% % \remind{More input for the acknowledgements?}
% \end{theacknowledgments}



\bibliographystyle{aipproc}   % if natbib is available

\bibliography{star,own}

\begin{thebibliography}{13}
\expandafter\ifx\csname natexlab\endcsname\relax\def\natexlab#1{#1}\fi
\providecommand{\enquote}[1]{``#1''}
\expandafter\ifx\csname url\endcsname\relax
  \def\url#1{\texttt{#1}}\fi
\expandafter\ifx\csname urlprefix\endcsname\relax\def\urlprefix{URL }\fi
\providecommand{\eprint}[2][]{\url{#2}}

\bibitem[{Baraffe} et~al.(1997)]{BCA1997}
I.~{Baraffe}, G.~{Chabrier}, F.~{Allard}, \& P.~H. {Hauschildt}, \emph{\aap}
  \textbf{327}, 1054--1069 (1997).

\bibitem[{Drechsel} et~al.(1995)]{DHL1995}
H.~{Drechsel}, S.~{Haas}, R.~{Lorenz}, \& S.~{Gayler}, \emph{\aap}
  \textbf{294}, 723--743 (1995).

\bibitem[{Ferguson} \& {James}(1994)]{FJ1994}
D.~H. {Ferguson}, \& T.~A. {James}, \emph{\apjs} \textbf{94}, 723--747 (1994).

\bibitem[{Ferguson} et~al.(1999)]{FLH1999}
D.~H. {Ferguson}, J.~{Liebert}, S.~{Haas}, R.~{Napiwotzki}, \& T.~A. {James},
  \emph{\apj} \textbf{518}, 866--872 (1999).

\bibitem[{Kurochkin}(1971)]{Kur1971}
N.~E. {Kurochkin}, \emph{Peremennye Zvezdy} \textbf{18}, 85--90 (1971).

\bibitem[{Liebert} et~al.(1995)]{LTN1995}
J.~{Liebert}, R.~W. {Tweedy}, R.~{Napiwotzki}, \& M.~S. {Fulbright},
  \emph{\apj} \textbf{441}, 424--428 (1995).

\bibitem[{Napiwotzki} et~al.(2003)]{NCD2003}
R.~{Napiwotzki}, N.~{Christlieb}, H.~{Drechsel}, H.-J. {Hagen}, U.~{Heber},
  D.~{Homeier}, C.~{Karl}, D.~{Koester}, B.~{Leibundgut}, T.~R. {Marsh},
  S.~{Moehler}, G.~{Nelemans}, E.-M. {Pauli}, D.~{Reimers}, A.~{Renzini}, \&
  L.~{Yungelson}, \emph{The Messenger} \textbf{112}, 25--30 (2003).

\bibitem[{P{\' e}quignot} \& {Tsamis}(2005)]{PT2005}
D.~{P{\' e}quignot}, \& Y.~G. {Tsamis}, \emph{\aap} \textbf{430}, 187--212
  (2005).

\bibitem[{Richer} et~al.(2003)]{RLS2003}
M.~G. {Richer}, J.~A. {L{\' o}pez}, W.~{Steffen}, G.~H. {Tovmassian},
  G.~{Stasi{\' n}ska}, \& J.~{Echevarr{\'{\i}}a}, \emph{\aap} \textbf{410},
  911--916 (2003).

\bibitem[{Stasi\'nska} et~al.(2005)]{STR2005}
G.~{Stasi\'nska}, {Tovmassian, G.~H.}, M.~G. {Richer}, M.~{Pe\~na},
  R.~{Napiwotzki}, C.~{Charbonnel}, \& L.~{Jamet}, \enquote{{The chemical
  composition of PN\,G135.9+55.9, the most oxygen planetary nebula},} in
  \emph{Proc.\ IAU Symp.\ 228, From Lithium to Uranium: Elemental Tracers of
  Early Cosmic Evolution}, edited by V.~Hill, P.~Francois, \& F.~Primas, 2005,
  p. in press.

\bibitem[{Tovmassian} et~al.(2001)]{TSC2001}
G.~H. {Tovmassian}, G.~{Stasi{\' n}ska}, V.~H. {Chavushyan}, S.~V. {Zharikov},
  C.~{Gutierrez}, \& F.~{Prada}, \emph{\aap} \textbf{370}, 456--467 (2001).

\bibitem[{Tovmassian} et~al.(2004)]{TNR2004}
G.~H. {Tovmassian}, R.~{Napiwotzki}, M.~G. {Richer}, G.~{Stasi{\' n}ska}, A.~W.
  {Fullerton}, \& T.~{Rauch}, \emph{\apj} \textbf{616}, 485--497 (2004).

\bibitem[{Wood} et~al.(1995)]{WRZ1995}
J.~H. {Wood}, E.~L. {Robinson}, \& E.-H. {Zhang}, \emph{\mnras} \textbf{277},
  87--94 (1995).

\end{thebibliography}

\end{document}